\documentclass[conference]{IEEEtran}
\usepackage{amsmath,bm,mathtools}
\usepackage{amssymb}
\usepackage{amsfonts}
\usepackage{graphicx}
\usepackage{epsfig}
\usepackage{subfigure}
\usepackage{psfrag}
\usepackage{xcolor}
\usepackage{citesort}
\usepackage{srcltx}
\usepackage{algorithm}
\title{Joint Task Assignment and Wireless Resource Allocation for Cooperative Mobile-Edge Computing}
\author{\IEEEauthorblockN{Hong Xing${}^\dag$, Liang Liu${}^\ddag$, Jie Xu{}$^\S$, and Arumugam Nallanathan${}^\dag$}\\
\IEEEauthorblockA{${}^\dag$School of EECS, Queen Mary University of London, London, U.K.\\
${}^\ddag$Department of ECE, University of Toronto, Toronto, Canada\\
${}^\S$School of Information Engineering, Guangdong University of Technology, Guangzhou, China\\
E-mails:~h.xing@qmul.ac.uk,~lianguot.liu@utoronto.ca,~jiexu@gdut.edu.cn,~a.nallanathan@qmul.ac.uk}}

\begin{document}
\maketitle

\begin{abstract}
This paper studies a multi-user cooperative mobile-edge computing (MEC) system, in which a local mobile user can offload intensive computation tasks to multiple nearby edge devices serving as helpers for remote execution. We focus on the scenario where the local user has a number of independent tasks that can be executed in parallel but cannot be further partitioned. We consider a time division multiple access (TDMA) communication protocol, in which the local user can offload computation tasks to the helpers and download  results from them over pre-scheduled time slots. Under this setup, we minimize the local user's computation latency by optimizing the task assignment jointly with the time and power allocations, subject to individual energy constraints at the local user and the helpers. However, the joint task assignment and wireless resource allocation problem is a mixed-integer non-linear program (MINLP) that is hard to solve optimally. To tackle this challenge, we first relax it into a convex problem, and then propose an efficient suboptimal solution based on the optimal solution to the relaxed convex problem. Finally, numerical results show that our proposed joint design significantly reduces the local user's computation latency, as compared against other benchmark schemes that design the task assignment separately from the offloading/downloading resource allocations and local execution.
%We aim to answer the question that how to optimally assign these independent tasks to different helpers for computation performance maximization, by taking into account the wireless channel conditions for offloading/downloading, as well as the computation and communication resource constraints at different nodes.

%Mobile edge computing бн.. Utilizing mobile edge computing, in this work we study a system consisting of one user with M tasks and N helpers, in which the user can opportunistically offload the tasks to the helps based on their communication and computational capacities. Under this model, we formulate a joint task assignment and wireless resource allocation problem to minimize the total computation time. This is a mixed-integer problem, which is in general NP hard. In this paper, we бн (describe your algorithm).  By simulation, we show that the proposed mobile edge computing based task assignment scheme can significantly reduce the total computation time as compared to the case when all the tasks are computed locally. Moreover, we also show that our proposed task assignment algorithm outperforms heuristic schemes, e.g., (only based on channel capacity or computational capacity). This shows that task assignment should be optimized based on both communication and computation.
\end{abstract}

\IEEEpeerreviewmaketitle

\newtheorem{definition}{\underline{Definition}}[section]
\newtheorem{fact}{Fact}
\newtheorem{assumption}{Assumption}
\newtheorem{theorem}{\underline{Theorem}}[section]
\newtheorem{lemma}{\underline{Lemma}}[section]
\newtheorem{corollary}{\underline{Corollary}}[section]
\newtheorem{proposition}{\underline{Proposition}}[section]
\newtheorem{example}{\underline{Example}}[section]
\newtheorem{remark}{\underline{Remark}}[section]
\newcommand{\mv}[1]{\mbox{\boldmath{$ #1 $}}}
\newcommand{\mb}[1]{\mathbb{#1}}
\newcommand{\Myfrac}[2]{\ensuremath{#1\mathord{\left/\right.\kern-\nulldelimiterspace}#2}}

\section{Introduction}
It is envisioned that by the year of 2020, around $50$ billions of interconnected Internet of Things (IoT) devices will surge in wireless networks, featuring new applications such as video stream analysis, augmented reality, and autonomous driving. The unprecedented growth of such applications demands intensive and latency-critical computation at these IoT devices, which, however, is hardly affordable by conventional mobile computing systems. To address such new challenges, mobile-edge computing (MEC) has been identified as a promising solution by providing cloud-like computing functions at the network edge  \cite{zhang2017FRAN,mao2017survey,kumar2013survey,ETSI14,cisco15}.
%As such, users (e.g., IoT devices) can offload their computation tasks to nearby APs and then download results from them after remote execution, which improves computation capability, enhances energy efficiency, and reduces latency.

%Technical standards for MEC are being developed by the European Telecommunications Standard Institute (ETSI) \cite{ETSI14}, while industrial innovations
%exemplified by, e.g., Cisco's characterization of fog computing \cite{cisco15}, and Microsoft's recently published machine learning algorithm tailored for IoT devices \cite{Microsoft17},
%are also dedicated to fulfilling technical specifications \cite{cisco15}.

%The energy beamformer, the CPU frequencies as well as the offloaded bits at each user, and the time allocations were jointly optimized to minimize the energy consumption at the AP.

MEC has received growing research interests in both academia and industry. To maximally reap the benefit of MEC, it is critical to jointly manage the radio and computation resources for performance optimization \cite{Barbarossa2014survey}. For instance, \cite{Yu2016OFDMA} investigated an MEC system with orthogonal frequency division multiple access (OFDMA)-based computation offloading, in which the subcarrier allocation for offloading and the users' central processing unit (CPU) frequencies for local computing were jointly optimized to minimize the energy consumption at mobile devices. \cite{You2017offloading} considered multi-user MEC systems with both time division multiple access (TDMA) and OFDMA-based offloading, in which the optimal resource allocation policies were developed by taking into account both wireless channel conditions and users' local computation capabilities.
Furthermore, a new multi-user MEC system was studied in \cite{Wang2017wksp} by exploiting multi-antenna non-orthogonal multiple access (NOMA)-based computation offloading. In addition, a wireless powered multi-user MEC system was developed for IoT systems in \cite{Wang2017MEC}, where the users conducted computation offloading relying on the harvested energy from a multi-antenna access point (AP) integrated with MEC servers.
In these prior works, the MEC servers are usually assumed to be of rich computation and energy resources, such that the computation time and/or the results downloading time are assumed negligible. This, however, may not be true in practice \cite{Osvaldo2017AR,Mao2017scheduling}, especially for scenarios where multiple lightweight edge devices such as cloudlets and smartphones are employed for cooperative mobile-edge computing.

On another front, in distributed computing systems, task assignment and task scheduling have been extensively studied to improve the computation quality of service (see, e.g., \cite{Alsalih2005energy-aware} and the references therein). For example, \cite{chen2017massiveD2D} studied the task assignment amongst multiple servers for parallel computation and \cite{Mao2017scheduling} investigated the scheduling of sequential tasks with proper order. However, this line of research often assumed static channel and computation conditions but ignored their dynamics and heterogeneity, thus making it difficult to be directly applied to MEC. Recently, there are few works considering joint task scheduling and communications resource management. For instance, \cite{Mao2017scheduling} jointly optimized the task scheduling and wireless power allocation in a single-user single-core MEC system, in which multiple independent computation tasks at the local user require to be sequentially executed at the MEC server.

In this paper, we study a multi-user cooperative MEC system, in which a local mobile user can offload a number of independent computation tasks to multiple nearby edge devices serving as helpers (such as smartphones, tablets, WiFi APs, and cellular base stations (BSs)) for remote execution. Assuming that the tasks can be executed in parallel but cannot be further partitioned, we consider a TDMA communication protocol, in which the local user offloads tasks and downloads computation results over pre-scheduled time slots. The contributions of this paper are summarized as follows. 1) We formulate the latency-minimization problem that jointly optimizes computation tasks assignment and time/power allocations for both tasks offloading and results downloading, subject to individual energy constraints at all the user and helpers. 2) Since the formulated problem is a mixed-integer non-linear program (MINLP) that is hard to solve optimally in general, we propose an efficient algorithm to obtain a suboptimal solution based on the optimal solution to a relaxed (convex) problem. 3) Simulation results show striking performance gain achieved by the proposed design in comparison with other benchmark schemes that design the task assignment separately from the offloading/downloading resource allocations and local execution.

The remainder of this paper is organized as follows. The system model is presented in Section \ref{sec:System Model}. The joint computation task assignment and time allocations problem is formulated in Section \ref{sec:Problem Formulation}. In Section \ref{sec:Proposed Joint task assignment and Time Allocations}, an effective joint optimization algorithm is proposed. Simulation results are provided in Section \ref{sec:Simulation Results}, with conclusion drawn in Section \ref{sec:Conclusion}.

\section{System Model}\label{sec:System Model}
We consider a multi-user cooperative MEC system that consists of one local mobile user, and $K$ nearby wireless edge devices serving as helper-nodes, denoted by the set \(\mathcal{K}=\{1,\ldots,K\}\), all equipped with single antenna. For convenience, we define the local user as the $(K+1)$-th node. Suppose that the local user has $L$ independent tasks to be executed, denoted by the set \(\mathcal{L}=\{1,\ldots,L\}\), and the input (output) data length of each task $l\in\mathcal{L}$ is denoted by \(T_l\) (\(R_l\)) in bits. In the considered MEC system, each task can be either computed locally, or offloaded to one of the $K$ helpers for remote execution. Let \(\mv\Pi\in\mathbb{R}^{L\times (K+1)}\) denote the task assignment matrix, whose $(l,k)$-th entry, denoted by \(\pi(l,k)\in\{0,1\}\), \(l\in\mathcal{L}\), \(k\in\mathcal{K}\cup\{K+1\}\), is given by
\begin{align*}
\pi(l,k)=\begin{cases}
1, & \mbox{if the}\ l\mbox{th task is assigned to the}\  k\mbox{th user,} \\
0, & \mbox{otherwise.}
\end{cases}
\end{align*}
Also, define $\mathcal{L}^{(k)}=\{l\in\mathcal{L}:\pi(l,k)=1\}$ as the set of tasks that are assigned to node $k$, $k\in\mathcal{K}\cup\{K+1\}$. At last, denote by \(C_{l,k}\) (in cycles per bit) the number of CPU cycles required for computing one input bit of the \(l\)th task at the $k$th node, $l\in\mathcal{L}$, $k\in\mathcal{K}\cup\{K+1\}$. Also denote the CPU frequency at the $k$th node as $f_k$ (in cycles per second), \(k\in\mathcal{K}\cup\{K+1\}\).

\begin{figure}[htp]
	\centering
	\includegraphics[width=3.3in]{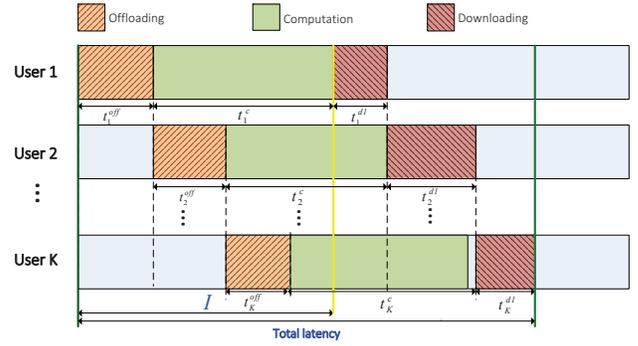}
	\caption{The TDMA-based frame for the proposed protocol.}\label{fig:frame protocol}
	%\vspace{-0.2in}
\end{figure}

%\begin{figure}[htp]
%	\centering
%	\subfigure[A multi-user MEC system consisting of one local mobile device and several nearby wireless facilities equipped with MEC servers. \label{subfig:system illustration}]{\includegraphics[width=2.8in]{eps/system_model.eps}}
%	\subfigure[The TDMA-based frame protocol.\label{subfig:frame protocol}]{\includegraphics[width=3.2in]{eps/frame_protocol.eps}}
%	\caption{The system model.}\label{fig:system model}
%	%	\vspace{-0.2in}
%\end{figure}
\subsection{Local Computing}
The tasks in the set \(\mathcal{L}^{(K+1)}\) are executed locally. Hence, the local computation time is given by
\begin{align}
t_0^c=
%\sum_{j=1}^{\sum_{l=1}^L\pi(l,K+1)C_lT_l}\Myfrac{1}{f_{0,j}}
\Myfrac{\sum_{l=1}^L\pi(l,K+1)C_{l,0}T_l}{f_0}.\label{eq:local computation time}
\end{align} The corresponding computation energy consumed by the local user is given by \cite{mao2017survey}
\begin{align}
E_0^c=
%\sum_{j=1}^{\sum_{l=1}^L\pi(l,K+1)C_{l,0}T_l}\kappa_0f_{0,j}^2
\kappa_0\sum_{l=1}^L\pi(l,K+1)C_{l,0}T_lf_0^2,\label{eq:local computation energy}
\end{align}
where $\kappa_0$ is a constant denoting the effective capacitance coefficient that is decided by the chip architecture of the local user.

\subsection{Remote Computing at Helpers}
On the other hand, the tasks in the set of \(\mathcal{L}^{(k)}\) requires to be offloaded to the $k$th node, \(k\in\mathcal{K}\), for remote execution. In this paper, we consider a three-phase TDMA communication protocol. As shown in Fig. \ref{fig:frame protocol}, the local user first offloads the tasks in the set $\mathcal{L}^{(k)}$ to the $k$th node, \(k\in\mathcal{K}\), via TDMA. Note that at each TDMA time slot, the local user merely offloads tasks to one helper. Then the helpers compute their assigned tasks and send the computation results back to the local user via TDMA. Similarly, at each time slot, there is merely one helper transmitting the results. In the following, we introduce the three-phase protocol in detail.
\subsubsection{Phase I: Task Offloading}
First, the tasks are offloaded to the helpers via TDMA. For simplicity, in this paper we assume that the local user offloads the tasks to the helpers with a fixed order of $1,2,\cdots,K$, as shown in Fig. \ref{fig:frame protocol}.
%In other words, the $0$th user offloads all the tasks in the set $\mathcal{L}^{1}$ to user $1$ first, and then all the tasks in the set $\mathcal{L}^{2}$ to user $2$, and so on.

Let $\bar{h}_k$ denote the channel power gain from the local user to the $k$th node for offloading, \(k\in\mathcal{K}\). The achievable rate from the local user to the $k$th node is given by (in bits/second)
\begin{align}
r_k^{off}=B\log_2\left(1+\frac{p_k^{off}\bar{h}_k}{\sigma_k^2}\right),\label{eq:offloading rate}
\end{align}where \(B\) in Hz denotes the available transmission bandwidth, \(p_k^{off}\) is the transmitting power at the local user for offloading tasks to the $k$th node, and $\sigma_k^2$ is the power of additive white Gaussian noise (AWGN) at the $k$th node. Then, the task offloading time for the $k$th node is given by\footnote{If no task is offloaded to node $k$, i.e., $\pi(l,k)=0$, $\forall l$, then the offloading rate \(r_k^{off}=0\), and we define $t_k^{off}=0$ and $p_k^{off}=0$ in this case.}
\begin{align}
t_k^{off}=\frac{\sum_{l=1}^L\pi(l,k)T_l}{r_k^{off}}. \label{eq:offloading time}
\end{align}
According to (\ref{eq:offloading rate}) and (\ref{eq:offloading time}), \(p_k^{off}\) is expressed as
\begin{align}
p_k^{off}=\frac{1}{h_k}f\left(\frac{\sum_{l=1}^L\pi(l,k)T_l}{t_k^{off}}\right), \label{eq:offloading power}
\end{align}where $h_k=\bar{h}_k/\sigma_k^2$ is the normalized channel power gain from the local user to the $k$th node, and \(f(x)\triangleq2^{\frac{x}{B}}-1\). The total energy consumed by the local user for offloading all the tasks to the helpers is then expressed as:
\begin{align}
E_0^{off}
=\sum_{k=1}^K\frac{1}{h_k}f\left(\frac{\sum_{l=1}^L\pi(l,k)T_l}{t_k^{off}}\right)t_k^{off}.\label{eq:offloading commun. energy}
\end{align}

\subsubsection{Phase II: Task Execution}
After receiving the assigned tasks in $\mathcal{L}^{(k)}$, the $k$th node proceeds with computing. Similar to (\ref{eq:local computation time}), the computation time of the $k$th node is given by
\begin{align}
t_k^c=\Myfrac{\sum_{l=1}^L\pi(l,k)C_{l,k}T_l}{f_k}. \label{eq:remote computation time}
\end{align} Its corresponding computational energy is thus given by
\begin{align}
E_k^c=\kappa_k\sum_{l=1}^L\pi(l,k)C_{l,k}T_lf_k^2,\label{eq:remote computation energy}
\end{align}where \(\kappa_k\) is the corresponding capacitance constant of the $k$th node.

\subsubsection{Phase III: Task Result Downloading}
After computing all the assigned tasks, the helpers begin transmitting computation results back to the local user via TDMA. Similar to the task offloading, we assume that the helpers transmit their respective computation results in the order of $1,\cdots,K$.
Let $\bar{g}_k$ denote the channel power gain from node $k$ to the local user for downloading. The achievable rate of downloading results from the $k$th node is then given by
\begin{align}
r_k^{dl}=B\log_2\left(1+\frac{p_k^{dl}\bar{g}_k}{\sigma_0^2}\right),\label{eq:feedback rate}
\end{align}where $p_k^{dl}$ denotes the transmitting power of the $k$th node, and $\sigma_0^2$ denotes the power of AWGN at the local user. The downloading time of the local user from the $k$th node is thus given by
\begin{align}
t_k^{dl}=\frac{\sum_{l=1}^L\pi(l,k)R_l}{r_k^{dl}}. \label{eq:downloading time}
\end{align}
Accordingly, the transmitting power of the $k$th node is expressed as
\begin{align}
p_k^{dl}=\frac{1}{g_k}f\left(\frac{\sum_{l=1}^L\pi(l,k)R_l}{t_k^{dl}}\right),\label{eq:downloading power}
\end{align}where $g_k=\bar{g}_k/\sigma_0^2$ denotes the normalized channel power gain from the $k$th node to the local user. The communication energy of the $k$th node for delivering its results to the local user is thus given by
\begin{align}
E_k^{dl}=\frac{1}{g_k}f\left(\frac{\sum_{l=1}^L\pi(l,k)R_l}{t_k^{dl}}\right)t_k^{dl}.\label{eq:downloading commun. energy}
\end{align}

Since TDMA is used in both Phase I and Phase III, each helper has to wait until it is scheduled. Specifically, the first scheduled helper, i.e., node $1$, can transmit its computation result to the local user only when the following two conditions are satisfied: first, its computation has been completed; and second, task offloading from the local user to all the $K$ helpers are completed such that the wireless channels begin available for data downloading, as shown in Fig. \ref{fig:frame protocol}. As a result, node $1$ starts transmitting its results after a period of waiting time given by
\begin{align}
I_1=\max\{ t_1^{off}+t_1^c,\sum_{k=1}^Kt_k^{off}\}, \label{eq:waiting time of helper 1}
\end{align} where \(t_1^c\) is the task execution time at node $1$ (c.f.~\eqref{eq:remote computation time}).

Moreover, for each of the other $K-1$ helpers, it can transmit the computation results to the local user only when: first, its computation has been completed; second, the $(k-1)$th node scheduled preceding to it has finished transmitting. Consequently, denoting the waiting time for node $k$ (\(k\ge 2\)) to start transmission as $I_k$, $I_k$ expressed as
\begin{align}
I_k=\max\{\sum_{j=1}^kt_j^{off}+t_k^c,I_{k-1}+t_{k-1}^{dl}\}. \label{eq:waiting time of helper k>=2}
\end{align} Accordingly, the completion time for all the results to finish being downloaded is expressed as
\begin{align}
T=I_K+t_K^{dl}. \label{eq:completion time}
\end{align}

To summarize, taking both local computing and remote execution into account, the total latency for all of the $L$ tasks to be executed is given as
\begin{align}
T^{\rm total}=\max\{t_0^c,T\}. \label{eq:total latency}
\end{align}

%It is easily verified that when the optimal $T^{\rm total}$ yields \(t_0^{c}>I_1+\sum_{k=1}^Kt_k^{dl}\), it is always possible, without loss of optimality, for one helper to slow down its transmission such that \(I_1+\sum_{k=1}^Kt_k^{dl}=t_0^{c}\) with its communication energy saved (c.f.~\eqref{eq:downloading commun. energy}). Therefore, $T^{\rm total}$ is simplified as \(T^{\rm total}=I_1+\sum_{k=1}^Kt_k^{dl}\) subject to \(t_0^{c}\le I_1+\sum_{k=1}^Kt_k^{dl}\).

\section{Problem Formulation}\label{sec:Problem Formulation}
In this paper, we aim at minimizing the total latency, i.e., \(T^{\rm total}\), by optimizing the task assignment strategy, i.e., $\pi(l,k)$'s, the transmission time for task offloading and result downloading, i.e., $t_k^{off}$'s and $t_k^{dl}$'s (equivalent to transmitting power as shown in (\ref{eq:offloading power}) and (\ref{eq:downloading power})), subject to the individual energy constraints for both the local user and the $K$ helpers as well as the task assignment constraints. Specifically, we are interested in the following problem:
\begin{subequations}
	\begin{align}
	\mathrm{(P1)}:\!\!&~\mathop{\mathtt{Minimize}}_{\mv\Pi,\{t_k^{off},t_k^{dl}\}}\!\!
	~T^{\rm total}\notag\\
	&\mathtt{Subject \ to}\notag\\
%	&\sum_{k=1}^Kt_k^{off}\le I_1,\label{C:the first constraint regarding I1}\\	
%	&t_1^{off}+\tfrac{\sum_{l=1}^L\pi(l,1)C_{l,1}T_l}{f_1^c}\le I_1,\label{C:the second constraint regarding I1}\\
	&E_0^c+E_0^{off}\le E_0,\label{C:energy constraint at the source}\\
	&E_k^c+E_k^{dl}\le E_k,\; \forall k\in\mathcal{K},\label{C:energy constraint at the kth helper}\\
	&\sum_{k=1}^{K+1}\pi(l,k)= 1, \; \forall l\in\mathcal{L},\label{C:task assignment constraint}\\
	&\pi(l,k)\in\{0,1\},\; \forall l\in\mathcal{L},\ k\in\mathcal{K}\cup\{K+1\},\label{C:binary constraint}\\
	&t_k^{off}\ge 0,\, t_k^{dl}\ge 0\; \forall k\in\mathcal{K}.\label{C:nonnegative time}
	\end{align}
\end{subequations}
The constraints given by
%\eqref{C:the first constraint regarding I1} and \eqref{C:the second constraint regarding I1} determine the waiting time before the local user starts downloading from node $1$ (c.f.~\eqref{eq:waiting time of helper 1});
\eqref{C:energy constraint at the source} and \eqref{C:energy constraint at the kth helper} represent the total energy constraints for the local user and the $k$th node, respectively; \eqref{C:task assignment constraint} guarantees that each task must be assigned to one node; and finally \eqref{C:binary constraint} ensures that each task cannot be partitioned.

%Problem (P1) is an MINLP and is in general NP-hard. It is worthy of noting that under given \(\mv \Pi\), (P1) proves to be convex with respect to (w.r.t.) \(t_k^{off}\)'s and \(t_k^{dl}\)'s, since the superposition of \(\max(\cdot)\) over linear functions turns out to be convex, and it can also be shown that \(E_0^{off}\) (c.f.~\eqref{eq:offloading commun. energy}) and \(E_k^{dl}\)'s (c.f.~\eqref{eq:downloading commun. energy}) are convex functions over \(t_k^{off}\)'s and \(t_k^{dl}\)'s, respectively. However, although the optimal solution to (P1) can be obtained by exhaustive search, it is computationally too expensive (as many as \((K+1)^L\) times of search) to implement in practice. Therefore, we propose in the following section a suboptimal scheme for (P1) to jointly optimize the task assignment and the transmission time/power.

\section{Proposed Joint Task Assignment and Time Allocations}\label{sec:Proposed Joint task assignment and Time Allocations}
The challenges in solving problem (P1) lie in two folds. First, the objective function (c.f.~\eqref{eq:completion time}) is a complicated function involving multiple \(\max\) functions due to recursive feature of $I_k$ (c.f.~\eqref{eq:waiting time of helper k>=2}), for \(k\ge 2\). Second, the task assignment variables are constrained to be binary (c.f.\eqref{C:binary constraint}). Hence, in this section we first simplify the objective function leveraging the structure of the optimal solution. Then for the equivalently transformed problem, we propose a suboptimal solution to deal with the binary constraints.

\subsection{Problem Reformulation}
First, the following lemma is required to simplify the objective function of (P1).
\begin{lemma}
The function \(h(y,t)=f\left(\frac{y}{t}\right)t\) monotonically decreases over \(t>0\).\label{lemma:decreasing function of the communications energy}
\end{lemma}
\begin{IEEEproof}
The monotonicity of the above function can be obtained by evaluating the first-order partial derivative of $h(x,t)$ with respect to (w.r.t.) $t$, and using the fact that \((1-x)e^x-1<0\), for \(x>0\).
\end{IEEEproof}
Then problem (P1) can be recast into an equivalent problem as stated in the following proposition.
\begin{proposition}
Problem (P1) is equivalent to the following problem:	
%Combining with the requirement of relaxing the constraints given by \eqref{C:binary constraint}, (P1) reduces to the following problem.
\begin{subequations}
	\begin{align}
	\mathrm{(P1\text{-}Eqv)}:\!\!&~\mathop{\mathtt{Minimize}}_{\mv\Pi,\{t_k^{off},t_k^{dl}\},I_1}\!\!
	~I_1+\sum_{k=1}^Kt_k^{dl}\notag\\
	&\mathtt{Subject \ to}\notag\\
	&\sum_{k=1}^Kt_k^{off}\le I_1,\label{C:the first constraint regarding I1}\\	
	&t_1^{off}+\tfrac{\sum_{l=1}^L\pi(l,1)C_{l,1}T_l}{f_1^c}\le I_1,\label{C:the second constraint regarding I1}\\
	&\tfrac{\sum_{l=1}^L\pi(l,K+1)C_{l,0}T_l}{f_0}\le I_1+\sum_{k=1}^Kt_k^{dl},\label{C:deadline constraint at the source}\\
	&\kern-0in\tfrac{\sum_{l=1}^L\pi(l,k)C_{l,k}T_l}{f_k^c}\le I_1+\sum_{j=1}^{k-1}t_j^{dl}-\sum_{j=1}^kt_j^{off},\notag\\
	&\forall k\in\mathcal{K}\setminus \{1\},\label{C:deadline constraint at the k>=2th helper}\\	
	&\eqref{C:energy constraint at the source}-\eqref{C:nonnegative time},
%	&\pi(l,k)\in[0,1],\; \forall l\in\mathcal{L},\ \forall k\in\mathcal{K}\cup\{K+1\},
	\end{align}
\end{subequations}
where the constraints given by \eqref{C:the first constraint regarding I1} and \eqref{C:the second constraint regarding I1} determine the waiting time of node $1$ (c.f.~\eqref{eq:waiting time of helper 1}); \eqref{C:deadline constraint at the source} follows by substituting \eqref{eq:local computation time} for \(t_0^c\) (c.f.~\eqref{eq:computation deadline constraint for the source}); and \eqref{C:deadline constraint at the k>=2th helper} are obtained by replacing \(t_k^c\)'s, \(k\ge 2\), with \eqref{eq:remote computation time} (c.f.~\eqref{eq:computation deadline constraint for helper k>=2}).
\end{proposition}

\begin{IEEEproof}
There are two possible cases for the optimal \(I_k\)'s given by \eqref{eq:waiting time of helper k>=2}: case 1) \(\sum_{j=1}^kt_j^{off}+t_k^c>I_{k-1}+t_{k-1}^{dl}\); and case 2) \(\sum_{j=1}^kt_j^{off}+t_k^c\le I_{k-1}+t_{k-1}^{dl}\). In line with Lemma~\ref{lemma:decreasing function of the communications energy}, the total transmitting energy of the $k$th node, i.e., \(E_k^{dl}\)'s (c.f.~\eqref{eq:downloading commun. energy}), monotonically decreases over \(t_k^{dl}\)'s. Hence, if the first case occurs, node $k-1$ (\(k\ge 2\)) can slow down its downloading, e.g., extending \(t_{k-1}^{dl}\), until \(I_{k-1}+t_{k-1}^{dl}=\sum_{j=1}^kt_j^{off}+t_k^c\), such that $I_k$ remains unchanged but the transmitting energy of node $k-1$ gets reduced. As such,  without loss of optimality, the two cases can be merged into one as
\begin{align}
I_k=I_{k-1}+t_{k-1}^{dl}, \; \forall k\in\mathcal{K}\setminus \{1\}, \label{eq:recursive I_k}
\end{align} subject to the computation deadline constraints given by
\begin{align}
t_k^c\le I_{k-1}+t_{k-1}^{dl}-\sum_{j=1}^kt_j^{off}, \; \forall k\in\mathcal{K}\setminus \{1\}. \label{eq:recursive computation deadline constraint for helper k>=2}
\end{align}
Since it follows from \eqref{eq:recursive I_k} that
\begin{align}
I_k=I_1+\sum_{j=1}^{k-1}t_j^{dl}, \label{eq:simplified waiting time of helper k>=2}
\end{align}
\eqref{eq:recursive computation deadline constraint for helper k>=2} reduces to
\begin{align}
t_k^c\le I_{1}+\sum_{j=1}^{k-1}t_j^{dl}-\sum_{j=1}^kt_j^{off}, \; \forall k\in\mathcal{K}\setminus \{1\}. \label{eq:computation deadline constraint for helper k>=2}
\end{align}
Furthermore, substituting \eqref{eq:simplified waiting time of helper k>=2} for $I_K$ in \eqref{eq:completion time}, $T$ is simplified as
\begin{align}
T=I_1+\sum_{k=1}^Kt_k^{dl}. \label{eq:simplified completion time}
\end{align} Then, plugging \eqref{eq:simplified completion time} into \eqref{eq:total latency}, the total latency given by \eqref{eq:total latency} turns out to be
\begin{align}
T^{\rm total}=\max\{t_0^c,I_1+\sum_{k=1}^Kt_k^{dl}\}. \label{eq:intermediate total latency}
\end{align}

On the other hand, it can be similarly verified that when the optimal \(T^{\rm total}\) given by \eqref{eq:intermediate total latency} yields \(t_0^{c}>I_1+\sum_{k=1}^Kt_k^{dl}\), it is always possible for one of the $K$ helpers to slow down its transmission with its communication energy saved such that \(I_1+\sum_{k=1}^Kt_k^{dl}=t_0^{c}\). Therefore, without loss of optimality, $T^{\rm total}$ can be further reduced to
\begin{align}
T^{\rm total}=I_1+\sum_{k=1}^Kt_k^{dl}, \label{eq:simplified total latency}
\end{align}
which is the objective function of Problem (P1-Eqv), and subject to
\begin{align}
t_0^{c}\le I_1+\sum_{k=1}^Kt_k^{dl}. \label{eq:computation deadline constraint for the source}
\end{align}
\end{IEEEproof}

It is also worthy of noting that to guarantee the feasibility of problem (P1) or (P1-Eqv), it is sufficient to have \(E_0>\sum_{l=1}^L(\kappa_0C_{l,0}T_lf_0^2+\ln2\sum_{k=1}^K\frac{1}{h_k}\frac{T_l}{B})\) and \(E_k>\sum_{l=1}^L(\kappa_kC_{l,k}T_lf_k^2+\frac{\ln2}{g_k}\frac{R_l}{B})\), \(\forall k\in\mathcal{K}\), which are assumed to be true throughout the paper, and thus we only focus on the feasible cases.

\subsection{Suboptimal Solution to (P1)}
Problem (P1-Eqv) is an MINLP and is in general NP-hard. Note that under given \(\mv \Pi\), (P1-Eqv) proves to be convex, since it is shown that \(E_0^{off}\) (c.f.~\eqref{eq:offloading commun. energy}) and \(E_k^{dl}\)'s (c.f.~\eqref{eq:downloading commun. energy}) are convex functions over \(t_k^{off}\)'s and \(t_k^{dl}\)'s, respectively. However, although the optimal solution to (P1-Eqv) can be obtained by exhaustive search, it is computationally too expensive (as many as \((K+1)^L\) times of search) to implement in practice. Therefore, in this subsection we propose a suboptimal solution to (P1-Eqv) (or (P1)) by jointly optimizing the task assignment and the transmission time/power.

We first replace the binary constraints given by \eqref{C:binary constraint} with the continuous ones given by
\begin{align}
&\pi(l,k)\in[0,1],\; \forall l\in\mathcal{L},\ \forall k\in\mathcal{K}\cup\{K+1\},\label{C:continous constraint}
\end{align}
which results in a relaxed problem denoted by (P1-Eqv-R). Since (P1-Eqv-R) proves to be jointly convex w.r.t. \(\mv\Pi\), \(\{t_k^{off},\\t_k^{dl}\}\), and \(I_1\), it can be efficiently solved by some off-the-shelf convex optimization tools such as CVX \cite{CVX}.

Next, denoting the optimal task assignment matrix \(\mv\Pi\) to (P1-Eqv-R) as \(\mv\Pi^\ast\), we propose to round off \(\pi^\ast_{l,k}\)'s to \(\hat\pi_{l,k}\)'s as follows such that \eqref{C:task assignment constraint} and \eqref{C:binary constraint} for (P1-Eqv) are satisfied.
\begin{align}
\hat\pi(l,k)=\begin{cases}
1,& \mbox{if}\ k=\hat k_l,\\
0,& \mbox{otherwise,}
\end{cases},\; \forall l\in\mathcal{L},\label{eq:suboptimal pi}
\end{align} where \(\hat k_l=\arg\kern-4pt\max\limits_{k\in\mathcal{K}\cup\{K+1\}}\kern-4pt\pi^\ast_{l,k}, \forall l\in\mathcal{L}\). As shown earlier, given \(\hat{\pi}_{l,k}\)'s as \eqref{eq:suboptimal pi}, (P1-Eqv) turns out to be jointly convex w.r.t. \(\{t_k^{off},t_k^{dl}\}\) and \(I_1\), and thus can again be efficiently solved by convex optimization tools to obtain optimal transmission time/power under given task assignment. The proposed algorithm for solving (P1) is summarized in Algorithm~\ref{alg:Algorithm I}.
\begin{algorithm}[ht]
\caption{Proposed Algorithm for Solving (P1)}\label{alg:Algorithm I}
\begin{enumerate}
\item Solve (P1-Eqv-R) using CVX;
\item Obtain the optimal task assignment matrix \(\mv\Pi^\ast\);
\item Round off \(\pi^\ast_{l,k}\)'s in accordance with \eqref{eq:suboptimal pi} yielding \(\hat{\mv\Pi}\);
\item Solve (P1-Eqv) given \(\hat{\mv\Pi}\) to obtain \(\{\hat t_k^{off},\hat t_k^{dl}\}\) and \(\hat I_1\).
\end{enumerate}
{\bf Output} the suboptimal solution to (P1) as \(\hat{\mv\Pi}\), \(\{\hat t_k^{off},\hat t_k^{dl}\}\), and \(\hat I_1\).
\end{algorithm}

\section{Simulation Results}\label{sec:Simulation Results}
In this section, we verify the effectiveness of the proposed joint task assignment and TDMA resource allocation against other baseline schemes. First, we provide two heuristic schemes: 1) `heuristic-$1$' assigns each task as per the channel gains only, i.e., \(k=\arg\min\limits_{k\in\mathcal{K}}\{\max\{\Myfrac{1}{h_k},\Myfrac{1}{g_k}\}\}\), \(\forall l\in\mathcal{L}\); and 2) `heuristic-$2$' assigns each task as per the computational time for executing this task, i.e., \(k_l=\arg\min\limits_{k\in\mathcal{K}}\Myfrac{C(l,k)T_l}{f_k}\), \(\forall l\in\mathcal{L}\). In addition, `random selection' solves (P1-Eqv) by randomly choosing a feasible \(\mv\Pi\). Moreover, since the theoretically optimal task assignment must be found by exhaustive search, we provide a near-optimal `random search' scheme that runs `random selection' for $1000$ times and selects the best solution. At last, in `local execution', the local user executes all the computation tasks locally.

The input data length $T_l$ is assumed to be uniformly distributed between \(0\) and \(10^4\), denoted by \(T_l\sim\mathcal{U}[0,10^4]\), \(\forall l\in\mathcal{L}\). Similarly, we set \(R_l\sim\mathcal{U}[0,10^3]\) and \(C_{l,k}\sim\mathcal{U}[0,10^3]\), \(\forall l\in\mathcal{L}\), \(k\in\mathcal{K}\). The $K$ helpers are located within a radius uniformly distributed within $0.5$km away from the local user. The wireless channel model consists of both  large-scale pathloss, and small-scale Rayleigh fading with an average channel power gain of $1$. The other parameters are set as follows unless otherwise specified: \(B=312.5\)~KHz, \(\sigma^2=-144\)~dB, \(\kappa=10^{-28}\) \cite{Wang2017MEC,You2017offloading}, \(K=5\), \(L=10\), \(f_0=1\)~GHz, \(f_k=2\)~GHz, and \(E_k=E_0=-20\)~dB, \(\forall k\in\mathcal{K}\).

\begin{figure}[htp]
	\centering
	\includegraphics[width=2.8in]{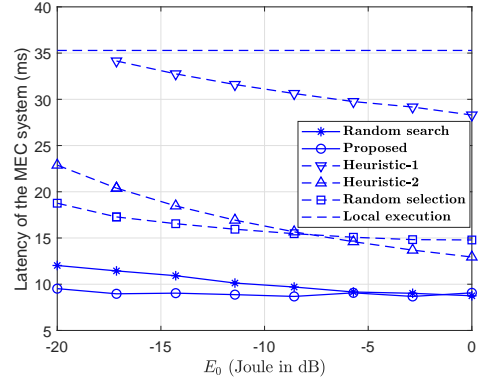}
	\caption{Total latency versus the energy constraints with \(E_k=E_0\), \(\forall k\in\mathcal{K}\).}\label{fig:latency_vs_energy}
\end{figure}

Fig. \ref{fig:latency_vs_energy} shows the total latency versus the energy constraints. It is observed that our proposed joint task assignment and time allocations outperforms all other schemes in most cases and keeps a negligible gap with the near-optimal `random search' under large energy constraints.
%This is because the proposed scheme is able to adaptively decide its task assignment as per the energy supply, e.g., distributing tasks to more users from $\mathcal{K}$ in low-energy case to prevent energy depletion on a single user, while only assigning tasks to users with competitive resources in high energy case.
It is also seen that `heuristic-2' only outperforms `random selection' when $E_0$ is larger than $-8.5$dB, since in the low-energy case, assigning tasks as per only computational resources may occur too much energy for computation and thus less for communications.

\begin{figure}[htp]
	\centering
	\includegraphics[width=2.8in]{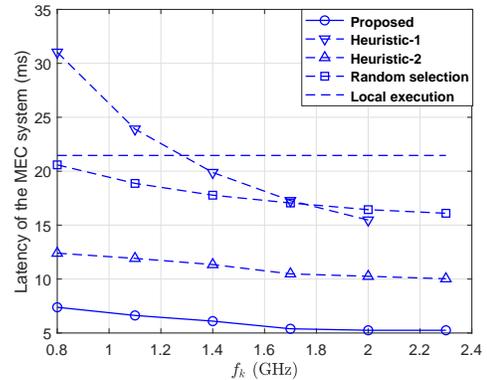}
	\caption{Total latency versus the helpers' CPU frequency with $f_0=1$GHz and $L=6$.}\label{fig:latency_vs_frequency}
\end{figure}
Fig. \ref{fig:latency_vs_frequency} compares the system latency versus the helpers' CPU frequency. It is seen that the total latency reduces with the helpers' frequency, which is intuitive. Moreover, both `heuristic-2' and the proposed scheme tend to be lower-bounded when \(f_k\)'s continues increasing, since under the same energy constraint, large \(f_k\)'s leads to significant computation energy expenditure, and thus the latency of the system is eventually bottlenecked by communications time. Furthermore, as `heuristic-1' only selects the helper with the best channel condition, all the tasks are then performed on this single helper, whose frequency thus needs to be sufficiently larger than the local frequency $f_0$ to surpass `local execution'. It is also worth noting that the performance of `heuristic-1' is not evaluated further when \(f_k\)'s is larger than $2$GHz, simply because operating with such high frequency violates its energy constraint.

\begin{figure}[htp]
	\centering
	\includegraphics[width=2.8in]{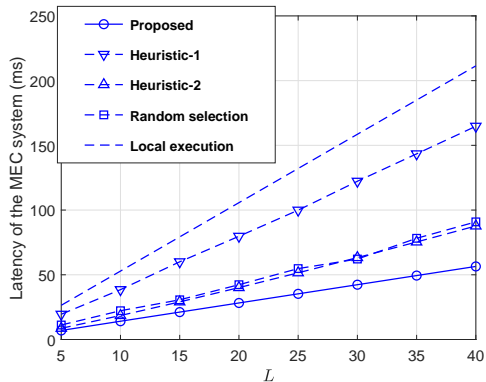}
	\caption{Total latency versus the number of computation tasks with $K=4$ and \(E_k=E_0=-10\)dB, \(\forall k\in\mathcal{K}\).}\label{fig:latency_vs_number_of_tasks}
\end{figure}
The impact of the total number of computation tasks on the latency is shown in Fig. \ref{fig:latency_vs_number_of_tasks}. With the number of tasks increasing, longer delay is expected for all schemes with the proposed design achieving the best performance, especially when $L$ becomes large. Unlike in Fig. \ref{fig:latency_vs_frequency}, 'heuristic-1' always outperforms 'local execution', since the communication-aware task assignment selects the helper with potentially short communications time to offload the tasks, while exploiting its high CPU frequency.

\section{Conclusion}\label{sec:Conclusion}

In this paper, we investigated joint task assignment as well as time and/or power allocations for a multi-user cooperative MEC system employing TDMA-based communications. We considered a practical task model where the local user has multiple independent computation tasks that can be executed in parallel. Under this setup, we aimed at minimizing the computation latency subject to individual energy constraints at the local user and the helpers. The latency minimization problem was formulated as an MINLP, which is difficult to be solved optimally. We proposed a low-complexity suboptimal scheme by first relaxing the integer variables (for task assignment) as continuous ones, then solving the relaxed problem, and finally constructing a suboptimal solution to the original problem based on the optimal solution to the relaxed problem. Finally, the effectiveness of the proposed scheme was verified by numerical results.
%\begin{appendices}
%\section{}\label{appendix:proof of rank-one for W_p^prime}
%Only an outline proof for Proposition~\ref{prop:rank one of W_p^prime} is given herein due to the space limitation. First, we present the Lagrangian function for problem \(\mathrm{(P3.1\text{-}2\text{-}SDR)}\). Next, apply Karush-Kuhn-Tucker (KKT) conditions to derive its first-order derivative condition w.r.t.~\(\mv W_p^\prime\), which leads to the final result of the proof.
%\end{appendices}

\bibliographystyle{IEEEtran}
\bibliography{MEC_ref}
\end{document}